\documentclass[12pt]{iopart}
\usepackage{graphicx}

\begin{document}

\title[
Multi-component fractional quantum Hall states in graphene
]{
Multi-component fractional quantum Hall states in graphene: SU(4) versus SU(2)
}

\author{C T\H oke$^{1,2}$ and J K Jain$^{3}$}
\address{$^{1}$Institute of Physics, University of P\'ecs, H-7624 P\'ecs, Hungary}
\address{$^{2}$BME-MTA Exotic Quantum Phases ``Lend\"ulet'' Research Group,
Budapest Univ. of Technology and Economics, Institute of Physics, Budafoki ut 8., H-1111 Budapest, Hungary}
\address{$^{3}$Department of Physics, 104 Davey Laboratory, Pennsylvania State University, University Park, Pennsylvania 16802, USA}
\ead{tcsaba@phy.bme.hu}

\begin{abstract} Because of the spin and Dirac-valley degrees of freedom, graphene allows the observation of one-, two- or four-component fractional quantum Hall effect in different parameter regions. We address the stability of various states in the SU(2) and SU(4) limits. In the SU(4) limit, we predict that new low-energy Goldstone modes determine the stability of the fractional quantum Hall states at 2/5, 3/7 etc.; SU(4) skyrmions are not found to be relevant for the low-energy physics. These results are discussed in light of experiments. 
\end{abstract}


\submitto{\JPCM}

\maketitle

\section{Introduction}

Experiments have raised the issue of the relative stability of various fractional quantum Hall states in graphene. The earliest results in suspended graphene in the two-terminal geometry \cite{Du} gave indications of fractional quantum Hall effect (FQHE) at $\nu=1/3$.  The more recent experiments by Dean \textit{et al} and Ghahari \textit{et al} \cite{fourterm} in four-terminal Hall bar geometry for graphene on boron nitride hexagonal substrate have revealed a much richer structure.  Specifically, they observe FQHE at filling factors $|\nu|=2/3, 4/3$ and  in the central ($n=0$) Landau level, and at $|\nu|=2+1/3, 2+2/3, 2+4/3, 2+5/3, 2+7/3$ in the $|n|=1$ Landau level. In addition, they also see evidence  FQHE at $|\nu|=1/3$ in the $n=0$ LL, but none at $|\nu|=2-1/3$. Hansel, Lafkioti, and V. Krsti\'c have also reported 4/3 FQHE in a four-terminal geometry \cite{fourterm2}. 
More recently, Feldman, Krauss, Smet and Yacoby \cite{fourterm3} have observed incompressible states in suspended flakes in the sequence $\nu = p/(2p \pm 1)$ for all $p \leq 4$ in the $|\nu|\le1$ range, but only at $\nu = 2-2/3, 2-2/5, 2-4/7$ and $2-4/9$ in the $1<|\nu|<2$ range.
The difference of the $0<|\nu|<1$ and $1<|\nu|<2$ interval, and the absence of FQHE at $2-1/3$ in particular, remains a puzzle.

Our treatment below builds upon our understanding of the FQHE in GaAs systems. In GaAs systems, the FQHE is associated with the formation of composite fermions (CFs), bound states of electrons and an even number of quantized vortices \cite{Jain89}. The electronic LL splits into Landau-like levels of composite fermions, called $\Lambda$ levels ($\Lambda$Ls), and the integer quantum Hall effect of composite fermions produces the prominently observed FQHE at $\nu=m/(2pm\pm 1)$, where $m$ is the number of completely filled $\Lambda$ levels and $2p$ is the number of vortices bound to electrons. The CF theory has been extended to include spin \cite{Wu,Park98}: with spin, FQHE occurs at the same fractions as before, namely $\nu=m/(2pm\pm 1)$, but now, in general, FQHE states with different spin polarizations are available at each fraction, analogous to the situation at integer fillings where different spin polarizations would occur depending on the strength of the Zeeman splitting relative to the cyclotron energy. Experiments have measured the transitions between differently polarized states, the actual spin polarizations of the states, CF $\Lambda$ level fan diagram, and the phase boundaries \cite{Duspin}, which are all found to be in good agreement with the CF theory. (For the integer quantum Hall effect, this physics is not relevant for GaAs, because the Zeeman energy is very small compared to the cyclotron energy, thus producing the state with the smallest spin polarization. However, for FQHE, the Zeeman energy can be of the order of the effective cyclotron energy of composite fermions, and therefore many different spin polarization occur.) We note that some of the states were written down in an early work of Halperin on multicomponent wave functions for the FQHE \cite{Halperin83}. 

The FQHE problem in graphene differs from that in GaAs in two respects. First, in graphene, each electron has four-components, because of two spin projections and two valleys, producing an approximate SU(4) symmetry when the Zeeman energy and the valley splittings are negligible.  Second, the linear dispersion leads to an interaction that is in general different from that in GaAs \cite{Yang,graphenesu2,su2more,graphenesu4,Goerbig06}. Our previous theoretical work on FQHE in graphene \cite{graphenesu2,graphenesu4,grapheneeven} has shown that composite fermions are formed in both $n=0$ and $|n|=1$ LLs. The FQHE is thus still predicted to occur at $\tilde\nu=m/(2pm\pm 1)$, where $\tilde\nu=\nu-(4n-2)$ is the fractional part of the filling factor in LL $n$, but, in general, many more kinds of incompressible states are now available at each fraction, depending on whether the system is in the SU(2) limit or SU(4). The actual ground state is determined by a competition between the CF kinetic energy, which favors occupation of the lowest available $\Lambda$ levels, the CF exchange energy, which favors occupation of $\Lambda$ levels with the same SU(4) index, and the Zeeman and valley splittings \cite{graphenesu4}.

Some of the observations on FQHE in graphene are 
consistent with theoretical expectations:  (i) The observed fractions have the form $m/(2pm\pm 1)$ as measured relative to a filled Landau level. (ii) The strength of FQHE in the $|n|=1$ LL is comparable to that in the $|n|=0$ LL, to be contrasted with GaAs systems \cite{graphenesu2,su2more}, where FQHE in the second LL is much weaker than that in the lowest LL. (iii) As predicted in \cite{Goerbig06,grapheneeven,Acta}, no FQHE has been observed at even denominator fractions in the $|n|=1$ LL.  Some other observations are unexpected, however. Noteworthy is the absence of fractions such as $2-1/3$ and $2-2/5$, while $2-2/3$ and $2-4/3$ are strong. Also puzzling is the appearance of FQHE at 1/3; one would have expected $2-1/3$ to be stronger given that its reference state at $\nu=2$ does not require any spontaneous symmetry breaking. In contrast, in the $|n|=1$ LL, $2+1/3$ and $2+2/3$ are both observed strongly. These fractions do not appear, at first sight, to be consistent with either SU(2) or SU(4) symmetry, thus posing a well defined puzzle whose resolution will shed important light on the nature of the FQHE in graphene.

Our aim in this paper is to determine the stability of various states by evaluating the gaps using the CF theory. The gaps in the SU(2) limit were evaluated previously for states of the form $\nu=m/(2pm+1)$ for the lowest LL of GaAs (these results carry over to the $n=0$ Landau level of graphene \cite{Davenport}) but not for the $n=1$ Landau level of graphene. In addition, gaps for the reverse-flux attached states at $\nu=p/(2pm-1)$ have not been calculated for technical reasons. (Lowest Landau level projection, which is needed to obtain states appropriate for the high field limit, is prohibitively expensive for large systems for the latter class of states.) However, the reverse-flux attached states are relevant to the graphene problem because the observed $2/3$ FQHE state is of that form \cite{Wu}. We estimate below the gaps for the states relevant to the recent experiments. In addition, we obtain the collective mode dispersions for various graphene FQHE states in both the SU(2) and SU(4) limits. 

The issue of the $1/3$ state in graphene has been addressed recently in exact diagonalization studies \cite{Papic}. Another approach for constructing FQHE states in the SU(4) limit can be found in \cite{GR}. A Chern-Simons treatment of composite fermions in the SU(4) limit is given in \cite{Modak}. 

Our paper is organized as follows.
Following necessary definitions in section \ref{model}, we study the excitation structure of the SU(2) limit in section \ref{su2limit} and the SU(4) limit in section \ref{su4limit}. We discuss our results in light of the recent experiments \cite{fourterm,fourterm2}.
Our conclusions are presented in section \ref{conclusion}.

\section{Model}
\label{model}

\subsection{Hamiltonian}

For electrons confined to a Landau level, the kinetic energy is quenched, and we need to consider the Hamiltonian that is diagonal in sublattice components (suppressed in our notation),
\begin{equation}
\label{hami}
\hat H=\frac{e^2}{\epsilon}\sum_{i<j}\frac{1}{|r_i-r_j|}-\sum_j {\Delta_Z\over 2}\hat\sigma_j^z+\sum_j {\Delta_V\over 2}\hat\tau_j^z.
\nonumber \end{equation}
Here $\epsilon$ is the dielectric constant of the background material, $\hat\sigma_i$ and $\hat\tau_i$ are Pauli matrices related to spin and Dirac valley indices, and $\Delta_Z$ and $\Delta_V$ are Zeeman and valley splittings.
For $\Delta_Z=0$ and $\Delta_V=0$, this Hamitonian possesses an SU(4) symmetry, with $\hat\sigma_z\otimes\mathbf 1,\mathbf 1\otimes\hat\tau_z,\hat\sigma_z\otimes\hat\tau_z$ (or any appropriate linear combinations thereof) defining the weights in SU(4) multiplets. Composite fermions acquire additional labels (quantum numbers) $\alpha^{(1)},\dots,\alpha^{(4)}$, which come from a basis chosen in the fundamental representation of SU(4), and can be chosen conveniently to be $\uparrow$$\uparrow$, $\uparrow$$\downarrow$, $\downarrow$$\uparrow$ and $\downarrow$$\downarrow$ where the first arrow refers to valley and the second to spin.  Let $(m_1,m_2,m_3,m_4)$ denote the incompressible ground state in which $m_1,\dots,m_4$ $\Lambda$ levels with label $\alpha^{(1)},\dots,\alpha^{(4)}$ are fully occupied. Which of these states is the ground state depends \cite{graphenesu4} on the values of $\Delta_Z$ and $\Delta_V$. A coupling between the spin and valley degrees of freedom and LL mixing are neglected in what follows.

We neglect the anisotropy of the interaction due to underlying
sublattice structure of the basis states, because such effects
are small, being proportional to the ratio of the lattice constant $a$ to the magnetic
length $\ell=\sqrt{\hbar c/e B}$. The study of the residual
lattice effects is beyond the scope of this paper.

Another important assumption of our model is the neglect of disorder. This is a standard assumption in most quantitative studies of FQHE, because it is not known how to include the effect of disorder in a reliable manner. It is possible that disorder might be affecting different FQHE states differently, thus affecting their relative stability. In the absence of a reliable quantitative treatment of disorder, while comparing our results with experiment we will make the simplest assumption that while the disorder causes a significant reduction of the excitation gap, it does not change the order of stability of various FQHE states. Strictly speaking, our results below ought to be considered as a prediction for future experiments in the limit of low disorder. 

The kinetic term still determines the Coulomb matrix elements via the single-particle orbitals when the Landau level is fixed.
For the $n=0$ LL the interelectron interaction is same as in GaAs, because in this LL the single-particle states are identical for carriers with linear and quadratic dispersions.
For $n=\pm1$, however, the 
two-component character of the single-particle states of massless Dirac fermions produces different interaction pseudopotentials than in GaAs.
Following standard practice, \cite{graphenesu2,graphenesu4,grapheneeven} we use a lowest LL basis with an effective interaction given in \cite{graphenesu4}, which reproduces the $|n|=1$ graphene pseudopotentials.  We employ the standard spherical geometry for our calculations \cite{Haldane},
in which electrons move on the surface of a sphere and a radial magnetic field is produced by a magnetic monopole of strength $Q$ (integer or half-integer) at the center, producing a magnetic flux $2Q(hc/e)$ through the surface.

\subsection{CF wave functions}

The general form of the CF wave functions in the spherical geometry is given by 
\begin{equation}
\Psi_{\nu}=\mathcal P_{\rm LLL}\Phi_{m}\prod_{j<k}(u_j v_k - v_j u_k)^{2p},
\label{cfwf0}
\nonumber
\end{equation}
Here, $u=\cos(\theta/2)e^{-i\phi /2}$ and $v=\sin(\theta/2)e^{i\phi /2}$. $\Psi_{\nu}$ is the wave function for interacting electrons at $\nu$ and $\Phi_{m}$ is the wave function for noninteracting electrons at filling factor $m$, with $\nu=m/(2pm\pm 1)$. In the spherical geometry, $\Psi_{m}$ is constructed at monopole strength $q$, producing $\Psi$ at monopole strength $Q=q+p(N-1)$. $\mathcal P_{\rm LLL}$ projects the state into the lowest LL \cite{JK}. The orbital part of the wave function has the same form as above even for the SU(2) and SU(4) generalizations, which we discuss below.

\subsection{Effective interaction}

The electron-electron interaction is conveniently described in terms of pseudopotentials \cite{Haldane} $V_m$, which give the energy of two electrons in relative angular momentum $m$.
As the orbitals in the lowest LL in graphene are identical to those in the conventional two-dimensional electron gas, the pseudopotentials are identical too.
The problem of interacting electrons in the $n=1$ LL of graphene can be mapped into a problem of interacting electrons in the lowest LL with the effective pseudopotentials \cite{su2more} 
\begin{equation}
\label{effective}
V_m^{(1)\textrm{gr.}}=\int\frac{d^2k}{(2\pi)^2}\frac{2\pi}{k}
\frac{1}{4}\left(L_1\left(\frac{k^2}{2}\right) + L_{0}\left(\frac{k^2}{2}\right)\right)^2
e^{-k^2}L_m(k^2),
\end{equation} 
where $L_n$ are Laguerre polynomials.

The Monte Carlo evaluation of the energy of variational wave functions, on the other hand, requires a real-space interaction.
To simplify calculation, we construct the wave functions in the lowest LL, and use an effective real-space potential corresponding to the effective pseudopotentials $V^{(1)\textrm{gr.}}_m$ in the lowest LL basis.
Following a well-tested procedure \cite{secondll,Park98}, we use the form
\begin{equation}
V^{\textnormal{eff}}(r)=\frac{1}{r}+\sum_{i=0}^M c_i r^i e^{-r},
\label{form}
\end{equation}
and fit the coefficients $c_i$ to reproduce the first $M+1$ pseudopotentials $V^{(1)\textrm{gr.}}_m$ of equation (\ref{effective}).
The coefficients $c_i$ obtained for $M=6$ are given in \cite{graphenesu4}.

\section{SU(2) analysis}
\label{su2limit}

To weigh the relative importance of the relevant parameters, it is appropriate to consider the ratio of the Zeeman splitting to some relevant energy scale in the problem.  We consider $\kappa=\Delta_Z/(0.02 e^2/\epsilon \ell)=0.05 g \epsilon \sqrt{B[T]}$, where the energy in the denominator is the Fermi energy of the single component CF Fermi sea and $g$ is the Land\'e factor for the background semiconductor material \cite{Park98,Duspin}.
(The quantity $\ell=\sqrt{\hbar c/eB}$ is the magnetic length. We could alternatively have chosen the gap of a prominent FQHE state.)
The FQHE states in GaAs are typically fully spin polarized for $B>5$ T, i.e., for ratio $\kappa > $0.5 \cite{Park98,Duspin}.  For graphene, this ratio is 1.8 (0.6) at $B=36$ T (4 T), suggesting that the spin degree of freedom may be frozen in the experiments of \cite{fourterm}, which are performed at rather large $B$. The valley symmetry breaking in graphene is on the order of $a/\ell$, where $a$ is the lattice spacing \cite{Goerbig06}.  At $B=4$ and 36 T, this ratio is $a/l\approx$ 0.02 and 0.06, respectively.  While this issue requires further investigation, it cannot be ruled out that the spin is fully polarized but the valley symmetry may not be broken \cite{Comment4}. We therefore first consider the situation where the spin degree of freedom is frozen but valley symmetry is intact. (The analysis of the SU(4) limit is given in the next section.)

In the SU(2) limit, the orbital part of the wave functions 
at the maximal weight states \cite{Hamermesh} of an SU(2) multiplet at monopole strength $Q$ are given by equation (\ref{cfwf0}), with 
\begin{equation}
\Phi_m=\Phi^{(1)}_{m_1}\Phi^{(2)}_{m_2}
\end{equation}
with $m_1+m_2=m$ and the superscript denotes the electron species. In other words, the state consists of $m_1$ and $m_2$ Landau bands of up and down pseudospins, respectively occupied. (Here, pseudospin refers to the valley SU(2) symmetry.) One can show that this wave function satisfies Fock's cyclic condition \cite{Hamermesh}, i.e., it is annihilated by any attempt at further antisymmetrization; this ensures that it is an eigenstate of the total pseudospin operator. The energies of variously valley-polarized states can be evaluated using standard methods, from which the phase diagram as a function of the valley Zeeman energy can be constructed \cite{Park98}.  The valley degree of freedom behaves very much like the spin degree of freedom. In fact, the predictions of the CF theory for two component systems have been verified semiquantitatively \cite{Shayegan} for another system with two valleys, namely AlAs quantum wells, where the valley splitting (analogous to the Zeeman energy) can be controlled by application of in-plane symmetry breaking strain.

\begin{figure}[!htbp]
\begin{center}
\includegraphics[width=\columnwidth,keepaspectratio]{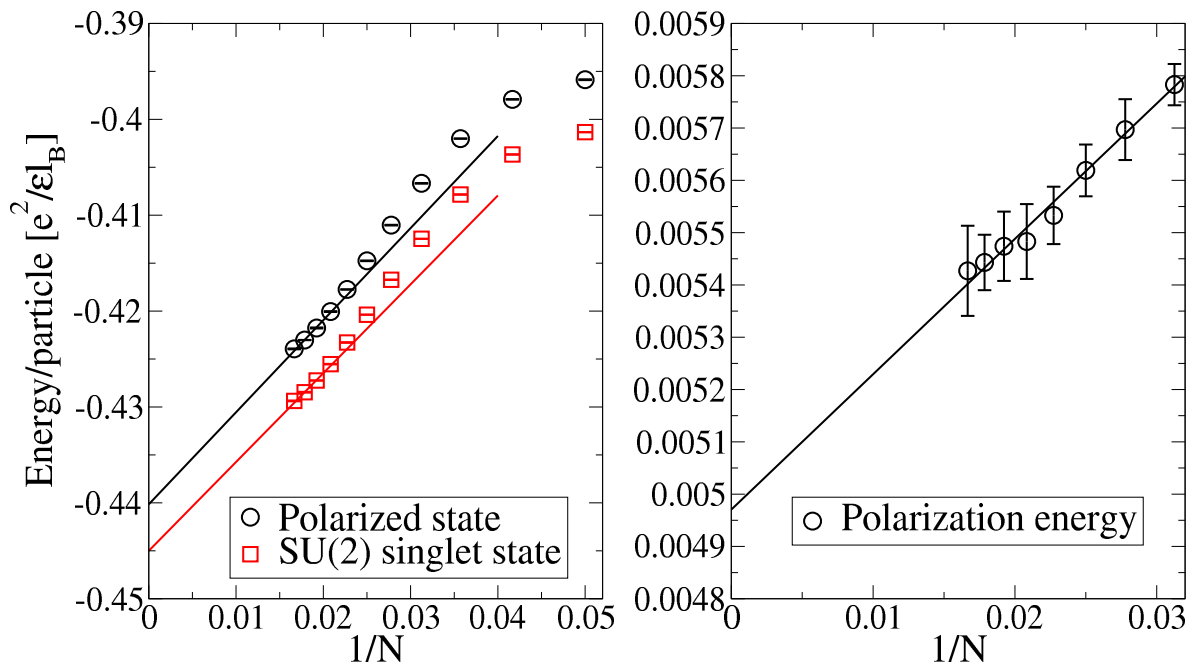}
\end{center}
\caption{\label{diff}
(a) The extrapolation of the energy of the fully polarized and the SU(2) singlet state at $\tilde{\nu}=2/5$ in the $|n|=1$ Landau level of graphene to the thermodynamic limit. $N$ is the number of particles. 
Only the leftmost four points, corresponding to $N\ge48$, are included in the fit because smaller systems show finite size effects.
(b) The polarization energy at $\tilde{\nu}=2/5$. The scaling is linear for $N\ge32$.
}
\end{figure}

With spin frozen, 1/3 is valley-polarized, whereas 2/3 and 2/5 are valley singlets.
(These states are related by particle-hole symmetry to 5/3, 8/5, and 4/3.)
That $\tilde{\nu}=2/5$ and $2/3$ are singlets in the $n=0$ Landau level follows from the analogous result established in conventional two-dimensional electron systems such as GaAs \cite{Park98,Wu}.

For the $\tilde{\nu}=2/5$ in the $|n|=1$ LL of graphene the energies of the 2/5 FQHE states in the thermodynamic limits can be obtained using the CF wave functions, and are given by $-0.4450(2) e^2/\epsilon\ell$ for the singlet and $-0.4401(2) e^2/\epsilon\ell$ for the polarized ground state energies, as shown in Fig.~\ref{diff}(a). The polarization energy, i.e., the energy difference between the polarized and the SU(2) singlet states, has been estimated to be $0.00497(9) e^2/\epsilon\ell$ in the thermodynamic limit as shown in Fig.~\ref{diff}(b). This confirms that, to the extent the wave functions are accurate,  the energy ordering of states predicted by the CF theory remains same in the $|n|=1$ LL as in the $n=0$ LL. We note that Shibata and Nomura \cite{valleysinglet} conclude, based on a density matrix renormalization group calculation, that the ground state here is {\em polarized} even in the absence of a symmetry breaking field; this conclusion, different from ours, is based on a study of smaller systems and without extrapolation to the thermodynamic limit.

We assume in what follows that the valley splitting is small but large enough to suppress the CF valley-skyrmions (CF skyrmions are rapidly suppressed by the Zeeman/valley splitting \cite{CFSkyrmion}), and the relevant excitation is a particle-hole pair of composite fermions.
If composite fermions were completely noninteracting, the 1/3 gap would be the smallest, as it does not involve CF cyclotron energy. The composite fermions, however, do feel a residual interaction, which necessitates a quantitative evaluation of the gaps, which can be accomplished using the standard methods of the CF theory \cite{JK,compdetails}. The energy gaps of the lowest charged excitations shown in Table \ref{gaps1} have several unexpected features \cite{previous}: The 2/3 and 4/3 states in the $n=1$ LL are the strongest, followed by 2/3 and 4/3 in the $n=0$ LL and 1/3 and 5/3 in the $n=1$ LL. Next most stable are 1/3 and 5/3 in the $n=0$ LL. 
These calculations are consistent with the observations that 2/3 and 4/3 are the strongest in the $n=0$ LL (producing FQHE at $2-2/3$ and $2-4/3$), both 1/3 and 2/3 are strong in the $n=1$ LL (resulting in FQHE at $2+1/3$, $2+2/3$, $4\pm 1/3$, $4\pm 2/3$), and 2/5 and 8/5 are not yet observed.

\begin{table}[htb]
\begin{center}
\begin{tabular}{c|c|c}
\hline\hline
$\tilde\nu$  & $n=0$ & $|n|=1$ \\
\hline
${1 \over 3}$ (${5 \over 3}$) & 0.073(1) & 0.095(2)\\
${2 \over 3}$ (${4 \over 3}$) & 0.11(4) & 0.21(14)\\
${2 \over 5}$ (${8 \over 5}$) & 0.0390(6) & 0.0475(9) \\
\hline\hline
\end{tabular}
\end{center}
\caption{\label{gaps1}
The smallest gaps for the creation of a far separated particle-hole pair of composite fermions.
The 1/3 ground state is fully spin and valley-polarized, whereas the 2/3 and 2/5 states are spin polarized but valley singlets.
Neglecting LL mixing, the gaps at $\nu$ and $2-\nu$ are equal. All energies are quoted in units of $e^2/\epsilon\ell$.
The 2/3 gaps come from a combination of exact diagonalization and CF theory; the rest comes from CF theory only.
}
\end{table}

While the observed order of stability is consistent with theory, the actual values of gaps are not. The largest observed gap \cite{fourterm} of $\sim16$ K at $\nu=4/3$ and $B=35$ T is much smaller than the theoretical estimate $\sim 0.1 e^2/\epsilon \ell\sim 110$ K (for $\epsilon=3$). 
The significant discrepancy is no doubt in large part due to disorder, and suggests that improvements in sample quality should reveal much further FQHE structure.

\section{SU(4) analysis}
\label{su4limit}

We next turn to the SU(4) limit, achieved when the valley and the spin states are all approximately degenerate and may be applicable to experiments at very low $B$. In this limit, if composite fermions were completely non-interacting, only FQHE states of the form $m/(2pm\pm 1)$ with $m=4, 8, \cdots$ would be observable, but as seen above, the residual interaction between composite fermions results in a spontaneous symmetry breaking, producing FQHE at all of these fractions \cite{graphenesu4}. We now show that new Goldstone modes become available, which are an SU(4) generalization of the spin waves of composite fermions considered previously in the context of GaAs FQHE \cite{Mandal}. (Similar mechanism has been considered for QHE at $\nu\neq 4n-2$ integer fillings in graphene \cite{Yang} and in bilayer systems \cite{bilayersu4}.) As a result of the presence of these modes, fractions such as 2/5 and 3/7 are expected to be much weaker than their SU(2) counterparts.

In the SU(4) generalization \cite{graphenesu4} of CF theory \cite{GR,Modak}, the orbital part of the ground state wave function at $\nu=m/(2pm\pm 1)$ for correlated electrons
in the maximal weight states \cite{Hamermesh} of an SU(4) multiplet at monopole strength $Q$ is still given by equation \ref{cfwf0}, but with
\begin{equation}
\Phi_m=\Phi_{m_1}^{(1)}\Phi_{m_2}^{(2)}\Phi_{m_3}^{(3)}\Phi_{m_4}^{(4)}
\end{equation}
where $\Phi_{m_s}^{(s)}$ is the Slater determinant wave function of $N_s$ electrons of the $\alpha^{(s)}$ species at monopole strength $q=Q-p(N-1)$, with $N=\sum_s N_s$ and $m=\sum_s m_s$. With $m_1\geq m_2\geq m_3\geq m_4$, it is easy to see that $\Psi$ satisfies Fock's cyclic condition \cite{Hamermesh}, i.e., it is annihilated by any attempt to antisymmetrize an electron $l$ of type $u$ ($\min_u\le l\le\max_u$) with respect to the electrons of type $t<u$,
\begin{equation}
\label{condi}
\left(1-\sum_{k=\min_t}^{\max_t}(k,l)\right)\Psi(\{z_j\})=0,
\nonumber
\end{equation}
where $(k,l)$ permutes indices $k$ and $l$.
For the CF-particle hole excitation, one must in general choose $\Phi$ to be an appropriate linear combination of Slater determinant states to ensure the Fock condition. 
In each SU(4) multiplet, represented by a Young tableau (YT)  $[M_1,M_2,M_3]$, where $M_i$ is the length of the $i$-th row ($M_i\ge M_{i+1}$ and the last $M_i$'s are not shown if zero), the maximal weight states can be constructed with the help of elementary group theory, and the eigenstates of the total orbital angular momentum $\hat L$ are obtained by standard Clebsch-Gordan expansion. We show the appropriate combinations pictorially for all single exciton excitations at 2/5, 3/7 and 4/9, along with their YT representations, in the subsequent figures. 
The interaction energies of the explicit wave functions are evaluated by Monte Carlo integration. For CF particle-hole excitation we identify their orbital angular momentum $L$ with their wave vector $k$ according to $k=L/\ell$, which yields the dispersion for the excitations, with the $k\to\infty$ limit giving the energy of a far separated CF particle-hole pair. All states in a given multiplet are degenerate for $\Delta_Z=0=\Delta_V$; for nonzero $\Delta_Z$ and $\Delta_V$ the energy splittings are straightforwardly determined. 

The CF ground state at $\tilde\nu=1/3$ has a single fully occupied $n=0$ $\Lambda$L.
As excitation to an unoccupied $n=0$ $\Lambda$L is already possible in SU(2) systems, we expect no new physics at this fraction.

\begin{figure*}[!htbp]
\begin{center}
\includegraphics[width=0.9\textwidth,keepaspectratio]{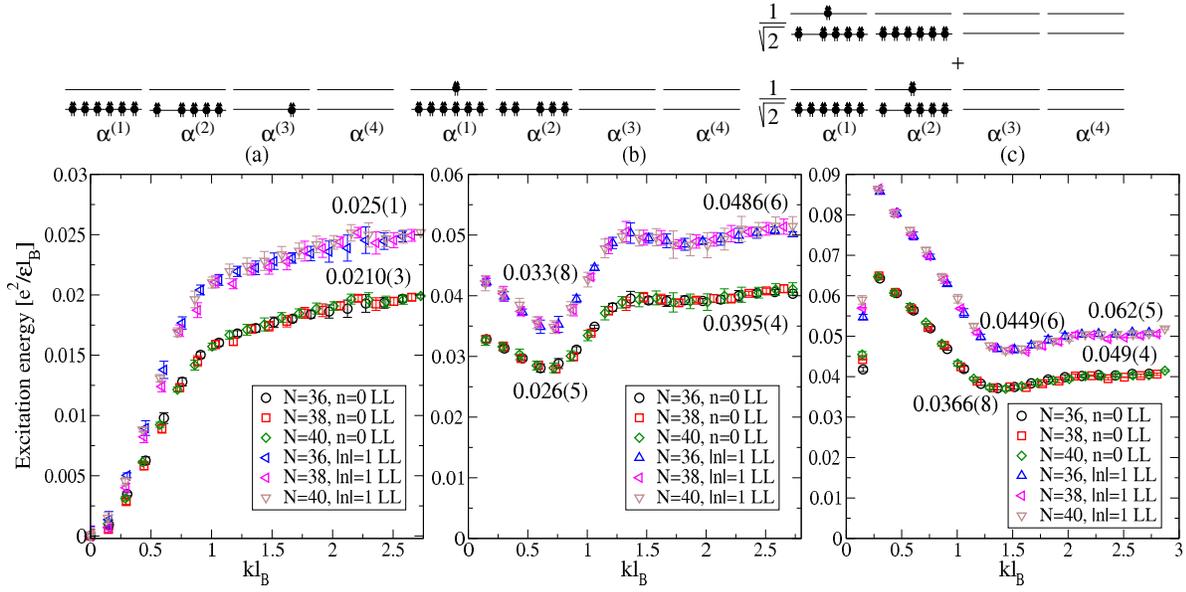}
\includegraphics[width=\textwidth,keepaspectratio]{twofifth2}
\end{center}
\caption{\label{twofifth}
Excitations and gaps \cite{compdetails} of the $\tilde\nu=2/5$ fractional quantum Hall effect in the $|n|=0,1$ graphene Landau levels.
The $\Lambda$ level structure for the maximum weight state in each mode is indicated in the diagrams at the top.
Mode (a), which has no analog in GaAs, determines the smallest charge gap.
The modes (b) and (c) correspond to singlet and triplet excitations in an SU(2) sector embedded in a SU(4) multiplet.
The transport and roton gaps shown near the largest momenta and the minimum, respectively, correspond to the thermodynamic limit.
The rotons in mode (b) destabilize the state when $\Delta_Z$ or $\Delta_V$ becomes greater than the roton energy. 
}
\end{figure*}

The $\tilde\nu=2/5$ state has two $n=0$ $\Lambda$Ls occupied in the ground state; assuming for definiteness that $\Delta_Z>\Delta_V$, the occupied $\Lambda$Ls are $\uparrow$$\uparrow$ and $\uparrow$$\downarrow$. It has two kinds of excitations depicted in Fig.~\ref{twofifth}: (i) The highly degenerate SU(4) spin wave, in which a CF is excited ``sideways" to an unoccupied $n=0$ $\Lambda$L, as shown in Fig.~\ref{twofifth}(a). This mode, represented by the YT $[N/2,N/2-1,1]$, is available only because of the SU(4) symmetry. (ii) Excitations within the SU(2) subspace of the ground state, in which a CF is raised ``up" to the second $\Lambda$L either flipping or preserving its valley index, represented, respectively, by YTs $[N/2+1,N/2-1]$ and $[N/2,N/2]$. The four cases give rise to a valley triplet (Fig.~\ref{twofifth}(b)) and a valley singlet (Fig.~\ref{twofifth}(c)), which are both gapped at all wavelengths. The triplet is split by the $\Delta_V$ and destabilizes the FQHE state when it reaches the ``valley-roton'' gap \cite{Mandal} of $\Delta^{\textnormal{roton}}_{2/5(b)}=0.026(5) e^2/\epsilon\ell$.
The  energy of the ``singlet-roton'' gap in mode (c) is unaffected by valley or Zeeman splittings. Transitions that raise a CF to the first excited level of an otherwise empty band are ignored, as these involve a loss of exchange energy and a kinetic energy cost at the same time and are likely to represent higher excitations.

Fig.~\ref{twofifth} shows that the smallest gap to creating a far-separated pair of charged excitations, which is the one relevant for transport experiments, corresponds to a CF exciton in mode (a). Because this energy is approximately half the gap in GaAs for either fully spin polarized or spin singlet 2/5 state \cite{Mandal}, the new Goldstone mode thus results in a substantial weakening of the 2/5 state in the SU(4) limit. From the SU(4) spin wave dispersion at small $k$ we estimate the energy of the SU(4) skyrmion, using the methods of \cite{Skyrmion}, to be 0.030(6) and 0.037(9) in the $n=0$ and $|n|=1$ Landau levels; this has a higher energy than the other excitations, so will not be relevant to either transport or spin polarization.

\begin{figure*}[!htbp]
\begin{center}
\includegraphics[width=0.9\textwidth,keepaspectratio]{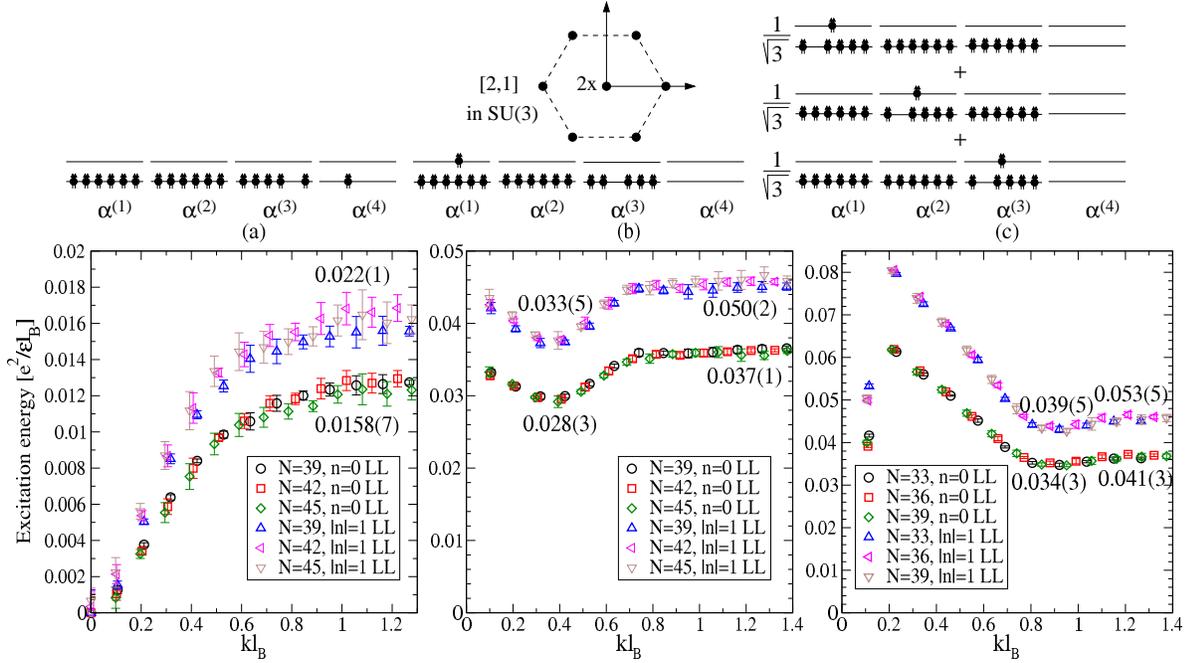}
\includegraphics[width=\textwidth,keepaspectratio]{threeseventh2}
\end{center}
\caption{\label{threeseventh}
Excitations and gaps \cite{compdetails} of the $\tilde\nu=3/7$ fractional quantum Hall effect in the $|n|=0,1$ graphene Landau levels. The spin-wave-like mode (a) produces the smallest charge gap. Modes (b) and (c) produce an SU(4) roton; the roton in (b) causes an instability with increasing  $\Delta_Z$ or $\Delta_V$. Roton and transport gaps are indicated.
2$\times$ denotes weight multiplicity.
}
\end{figure*}

At $\tilde\nu=3/7$ there exist ``sideways" excitations (Fig.~\ref{threeseventh}(a)) as well as those that stay within the SU(3) sector of the ground state (Fig.~\ref{threeseventh} (b) and (c)).
The nine possible excitons within the SU(3) sector can be reduced to a $[2,1]$ octet of SU(3) (see illustration above Fig.~\ref{threeseventh}(b)) and an SU(3) singlet.
The latter SU(3) multiplets are embedded in $[N/3+1,2,1]$ and $[N/3]$ SU(4) multiplets, respectively, and are all gapped. The SU(4) spin wave mode of Fig.~\ref{threeseventh}(a) is not gapped in the $k\to0$ limit and has the smallest energy at all wave vectors. The rather small gap to creating a far separated CF particle-hole pair in mode (a) makes the 3/7 state very delicate in the SU(4) limit.  The stiffness of the SU(4) spin wave in (a) could not be determined with sufficient accuracy, precluding an estimate of the SU(4) skyrmion energy at 3/7.

Weak symmetry breaking fields $\Delta_Z$ and $\Delta_V$ split the multiplet and select a state of reduced symmetry out of the SU(2)$\times$SU(2) decomposition \cite{Quesne} of the SU(4) multiplet for both the ground state and the exciton.
To illustrate we consider $\tilde{\nu}=2/5$ with $\Delta_Z>\Delta_V$. The spin and pseudospin (valley) quantum numbers are $(S,P)=(N/2,0)$ for the ground state and $(N/2-1,1\textnormal{ or }0)$, $(N/2,1)$ and $(N/2,0)$, respectively, for modes (a) to (c). 
Thus the fourfold degenerate mode (a) is raised by $\Delta_Z$ and split by $-\Delta_V,0,0,\Delta_V$; mode (b) is split similarly but it is not shifted; mode (c) is neither split nor shifted. For each mode as well as the ground state there are other copies displaced to higher energies in multiples of $\Delta_Z$, which in turn must merge into the continuum of other excitations; the structure of these higher copies follows from the SU(4) to SU(2)$\times$SU(2) decomposition \cite{Quesne}.

\begin{figure}[!htbp]
\begin{flushright}
\includegraphics[width=0.9\columnwidth,keepaspectratio]{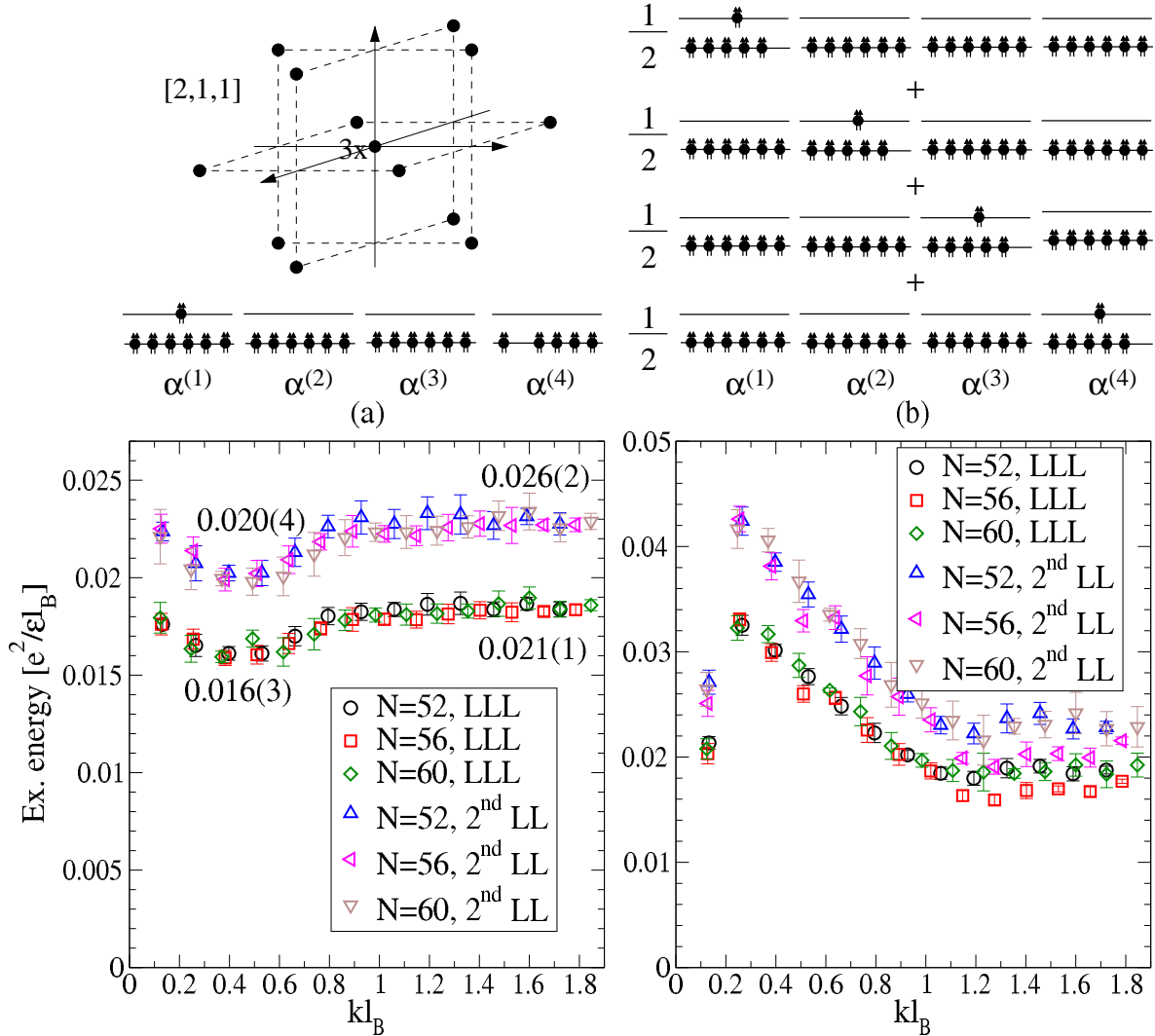}
\includegraphics[width=\columnwidth,keepaspectratio]{fourninth2}
\end{flushright}
\caption{\label{fourninth}
Excitations of the SU(4) singlet $\tilde\nu=4/9$ FQHE in the $|n|=0,1$ Landau levels.
The weight structure of the 15-fold degenerate mode (a) is shown.
Notice the roton minimum in mode (a), which is destabilizing if $\Delta_Z+\Delta_V\sim\Delta^{\textnormal{sf-roton}}_{4/9}$.
3$\times$ denotes weight multiplicity. Roton and transport gaps in mode (a) are indicated.
}
\end{figure}

We have also studied $\tilde{\nu}=4/9$, where there is no spin wave but the sixteen possible excitations into the second $\Lambda$L decouple into an SU(4) singlet (b) and the fifteen-fold degenerate mode (a) labeled by the YT $[2,1,1]$, see Fig.~\ref{fourninth}.
The large $k$ limits of lowest energy are 0.021(1) and 0.026(2) for the $n=0$ and $n=1$ LLs.
As a function of $\Delta_Z$ or $\Delta_V$ there are several possible transitions into CF states with other spin and valley quantum numbers \cite{graphenesu4}.
The roton and transport gaps in mode (b) could not be obtained with sufficient accuracy in the theormodynamic limit;
but in any finite systems they are not smaller than in mode (a). It is uncertain if mode (b) truly develops a roton gap.
Rotons in mode (a) destabilize the state at sufficiently high Zeeman energy.

\section{Conclusion}
\label{conclusion}

To conclude, we have argued that the spin degree of freedom may be frozen at high fields in graphene, and that many features of the 
FQHE in graphene at high magnetic field are consistent with SU(2) CF theory. In particular, quantitative calculations are consistent with the observed order of stability of states in $n=0$ and $|n|=1$ LLs, although the observed gaps are much smaller than the theoretical values. We have further shown that in the SU(4) limit, the FQHE states at $\tilde\nu=2/5$ and $3/7$ have a greatly reduced charge gap due to new SU(4) spin wave mode, which implies more stringent conditions for their observation.

We are thankful to Eva Andrei, Philip Kim, Cory Dean and Andrea Young for stimulating discussions. J.K.J. was supported in part by the NSF under grant no. DMR-1005536.  C. T. was supported by Science, Please! Innovative Research Teams, SROP-4.2.2/08/1/2008-0011 and the Bolyai Fellowship of the Hungarian Academy Sciences.
The authors acknowledge Research Computing and Cyberinfrastructure, a unit of Information Technology Services at The Pennsylvania State University,
and the National Information Infrastructure Development Program (Hungary) for providing high-performance computing resources and services.
We thank L\'aszl\'o Udvardi for assistance with the high-performance cluster at the Budapest University of Technology and Economics.


\begin{thebibliography}{99}
\bibitem{compdetails} The gaps shown are thermodynamic extrapolations from $N=10-30$ for 1/3, $N=20-40$ for 2/5, $N=24-45$ for 3/7, and $N=8-12$ for 2/3 (which requires reverse flux attachment).
\bibitem{previous} The gaps for 1/3 and 2/5 in the $n=0$ LL were reported previously in \cite{Mandal}. The gap values here are somewhat larger, but agree within statistical error.
\end{thebibliography}
\end{document}